\documentclass{aa}  
 
\usepackage{graphicx}
\usepackage{amsmath}
\usepackage{psfrag}
\usepackage{natbib}
\usepackage{color}
\bibpunct{(}{)}{;}{a}{}{,} 
\usepackage[dvips]{hyperref}
\usepackage{wasysym}
\newcommand{\beq}{\begin{equation}}
\newcommand{\eeq}{\end{equation}} 
\def\Div{\mathop{\hbox{div}}\nolimits}
\newcommand{\vu}{\vec{u}}
\renewcommand{\na}{ \vec{\nabla} }
\newcommand{\ltex}[1]{\quad \hbox{#1} \quad}
\newcommand{\disp}[1]{\displaystyle #1}
\newcommand{\vg}{\vec{g}}
\newcommand{\lc}{ \left[}
\newcommand{\rc}{ \right]}
\newcommand{\dnt}[1]{\frac{d  #1}{dt}}
\newcommand{\lp}{ \left(}
\newcommand{\rp}{ \right)}
\newcommand{\od}[1]{\mbox{${\cal O}(#1)$}}
\newcommand{\dnz}[1]{\frac{d  #1}{dz}}
\newcommand{\vpsi}{\vec{\psi}}
\newcommand{\ddnz}[1]{\frac{d^2  #1}{dz^2}}
\newcommand{\vf}{\vec{f}}
\newcommand{\KT}{{\cal K}_T}

\newcommand{\Amp}{{\cal A}}
\newcommand{\Fbot}{F_{\hbox{\scriptsize bot}}}

\newcommand{\Tbump}{T_{\hbox{\scriptsize bump}}}
\newcommand{\Kmin}{K_{\hbox{\scriptsize min}}}
\newcommand{\Kmax}{K_{\hbox{\scriptsize max}}}
\newcommand{\Ttr}{T_{\hbox{\scriptsize TR}}}
\newcommand{\Tion}{T_{\hbox{\scriptsize ionisation}}}
\newcommand{\rhotop}{\rho_{\hbox{\scriptsize top}}}
\newcommand{\Ttop}{T_{\hbox{\scriptsize top}}}
\newcommand{\dz}{\hbox{d}z}
\newcommand{\dV}{\hbox{d}V}

\begin{document}

\title{Direct numerical simulations of the $\kappa$-mechanism}

\subtitle{I.\ Radial modes in the purely radiative case}

\author{T.\ Gastine \and B.\ Dintrans}
\institute{Observatoire Midi-Pyr\'en\'ees, UMR5572, CNRS et Universit\'e
de Toulouse, 14 avenue Edouard Belin, 31400 Toulouse, France}

\offprints{thomas.gastine@ast.obs-mip.fr}

\date{\today,~ $Revision: 1.149 $, accepted for publication in A\&A}

\abstract
{Hydrodynamical model of the $\kappa$-mechanism in a purely radiative case.}
{First, to determine the physical conditions propitious to
$\kappa$-mechanism in a layer with a configurable conductivity hollow
and second, to perform the (nonlinear) direct numerical simulations
(DNS) from the most favourable setups.}
{A linear stability analysis applied to radial modes using a spectral
solver and DNS thanks to a high-order finite difference code are compared.}
{Changing the hollow properties (location and shape) lead to well-defined
instability strips. For a given position in the layer, the amplitude
and width of the hollow appear to be the key parameters to get unstable
modes driven by $\kappa$-mechanism. The DNS achieved from these more
auspicious configurations confirm the growth rates as well as structures
of linearly unstable modes. The nonlinear saturation follows through
intricate couplings between the excited fundamental mode and higher
damped overtones.} 
{}


\maketitle

\section{Introduction}

Since the beginning of the studies concerning Cepheids, it is well
known that convection occurs in Cepheids' envelopes, and thus changes
pulsation properties \citep[e.g. the reviews of][]{Gautschy-Saio,buchler-cepheid-review}.
The coldest ones, which are located next to the red edge of the
instability strip, have the more extended surface convective zones.
However, during many years Cepheids' oscillations models have used
the so-called ``frozen-in convection'' approximation which claims
that convective flux perturbations are negligible \citep{Baker1962}.
Such kind of models well predict the blue edge of the instability
strip but fail to explain the red edge as in this case, the strong
existing couplings between the convection and the oscillations are
not taken into account. This discrepancy becomes obvious with the
accumulation of accurate observations which show a narrower instability
strip than the theoretical one, i.e. modes which are linearly
unstable in the models are located outside the observational
strip \citep[][hereafter YKB98]{YKB}.

The main theoretical difficulty comes from the fact that convection
plays a crucial role on the pulsations while we know that convection
itself remains roughly described by mixing-length theories
\citep{BV2,BV1}. However, several attempts in the direction of time
dependent convection models (TDC) have been developed
\citep[e.g.][]{Unno,Gough,St}. Recent studies \citep[YKB98;][]{Bono}
rely on Stellingwerf's convection model \citep{St} or very similar
newer approaches to compute linear and nonlinear time evolution of
amplitudes of modes \citep{K,GW,WF}. The major problem of TDC is the
choice of the many free parameters introduced by the convection
model\footnote{e.g. the seven coefficients $[\alpha_c,\ \alpha_t,\
\alpha_{\nu},\ \alpha_{\Lambda},\ \alpha_s,\ \alpha_d$,\ $\alpha_p]$
in YKB98 or the eight ones in \cite{Koll02}.}. As these parameters are
not theoretically well determined, one should adjust them by fitting
the observations.

Another way to study the convection-pulsation interaction is to
achieve (nonlinear) direct numerical simulations (DNS) of the whole
hydrodynamical problem. The final aim of our work is to realise
such kind of simulations in 2-D and 3-D where a convective zone will be
coupled with a radiative one and unstable radial acoustic modes will
be self-consistently excited by $\kappa$-mechanism. However, as DNS
are highly time consuming, it is necessary to get in a first step
the appropriate initial conditions. That is why we have tried to
determine in this paper the physical conditions propitious to an
excitation based on $\kappa$-mechanism.

\citet{Eddington}, and then \citet{Zhevakin53} and \citet{1958Cox} have
introduced a mechanism linked to the opacity in ionisation regions,
the $\kappa$-mechanism, where $\kappa$ denotes opacity \citep[see
also the review of][]{Zhevakin}. They have shown that Cepheids' radial
oscillations are driven by a thermal heat engine as radial pulsations
have to be maintained thanks to a sustained physical process. That is
why, they imagined a Carnot-like thermodynamic cycle which stores heat
during compression phases while releasing it during the decompression
ones. This mechanism, now called the \textit{Eddington's valve}, can
only occur in specific regions of a star where the opacity varies so
as to block the radiative flux during compression phases. Now, opacity
tables such as the OPAL ones show strong increases in ionisation regions
of main (i.e. Hydrogen and Helium) or heavy elements that are
commonly called ``ionisation bumps'' \citep[e.g.][]{Seaton-Badnell}. As a
consequence these ionisation zones are locally responsible for modes
amplification. Moreover, beyond this criterion on opacity, these
ionisation regions have to be located in a very precise region of a star.

Indeed, if they are too deep or too close to the surface, the
driving they cause can be balanced by the damping occurring in other
regions. Therefore, an efficient ionisation region has to be located in
a specific place, called the \textit{transition region}, which marks
the shift from a quasi-adiabatic interior to a strongly non-adiabatic
surface. In classical Cepheids oscillating on the radial fundamental
mode, the overlap of this transition region with the ionisation one is
around $40\ 000$ K, which corresponds to the temperature of the helium
second ionisation \citep{Baker-Kippenhahn}. For first overtone
Cepheids, things are more intricate as one must take into account first,
the respective position between this HeII ionisation region and the nodal
line and second, the HeI/H region which also contributes to the driving
\citep{Bono}.

However, the location of these opacity bumps are not solely
responsible of the acoustic mode destabilisation. A careful treatment
of the $\kappa$-mechanism in Cepheid stars would also involve the
possible dynamical couplings with convective zones or, say, metallicity
effects through a realistic equation of state and opacity tables. The
corresponding physics is actually not fully understood from a theoretical
point of view and the purpose of our work is to sufficiently simplify
the hydrodynamical approach while keeping at the same time the
\textit{leading order} phenomenon responsible of the instability,
that is, the opacity bumps location.

That's for why we adopt a fully radiative layer of a perfect gas in
which a ionisation region is represented by a ``hollow'' in radiative
conductivity, corresponding to a ``bump'' in opacity. Strictly speaking,
the layer stability will depend on both temperature and density
variations, as the radiative conductivity is a function of these
two physical quantities. However, as the opacity strongly depends on
temperature, this state variable mainly controls the instability. As our
final aim is to realise an hydrodynamical study of $\kappa$-mechanism,
we will therefore neglect any dependence of radiative conductivity
on density and thus, our conductivity profile is merely a function
of temperature. The inferred advantage is to allow easy changes in
the different parameters of the ionisation region by setting both
position and shape of the conductivity hollow (i.e. its slope, width and
amplitude). As a consequence, it is possible to investigate a complete
parametric study of $\kappa$-mechanism in order to determine precisely
the physical conditions required by the instability.

We first consider the radiative and hydrostatic equilibria of our layer
with an appropriate conductivity profile. By adjusting both the
density value at the top of the layer and the flux at its bottom, we then
obtain a transition region in the middle of the computational domain.
Secondly, we investigate the linear stability analysis by solving
perturbation equations for the oscillations. We thus obtain the whole
spectrum, and can sort out unstable modes from stable ones. Therefore,
we are able to check the relevance of the transition region concept by
making every parameters of the conductivity hollow vary.

The first main result of our linear stability study is the confirmation
of the underlying conditions defining the transition region. With
different parametric studies, we obtain instability strips corresponding
to the fundamental mode, as both a minimum hollow width and amplitude
are needed to obtain unstable modes. These results are interpreted thanks
to the work integral to exactly determine the location of the driving
zone. As a consequence, this parametric study of $\kappa$-mechanism
provides us the physical quantities responsible for the instability.

Secondly, these auspicious conditions needed to drive the fundamental
mode constitute the starting point of 2-D DNS. Indeed, we check the
growth rates as well as the structure of the linearly unstable modes by
performing direct simulations until the nonlinear saturation of modes.

In \S\ref{hydro_eqs}, we first introduce the general oscillation
equations while the different conditions leading to the instability
are determined in \S\ref{ref:inst-cond}. In \S\ref{sec:hydro_model},
we present our hydrodynamical model. Linear stability analysis and
DNS results are thus given and compared in \S\ref{sec:results}
and \S\ref{sec:DNS}, respectively, before concluding in
\S\ref{sec:conclusion}.

\section{The general oscillations equations}
\label{hydro_eqs}

Our system is composed by a 2-D plane parallel layer of width $d$.
$\vec{z}$ denotes a cartesian coordinate, pointing upward on the
contrary of the constant gravity field $\vec{g}$. The gas is assumed
to be monatomic and perfect, so its equation of state is given by
\beq
p=R_*\rho T \ltex{and} \gamma\equiv \dfrac{c_p}{c_v}={5 \over 3},
\label{EOS}
\eeq
where $p$, $\rho$ and $T$ respectively denote pressure, density and
temperature, $R_*$ is the ideal gas constant and $\gamma$ the ratio
of specific heats $c_p$ and $c_v$.

\begin{figure}[htbp]
 \centering
 \includegraphics[width=7 cm]{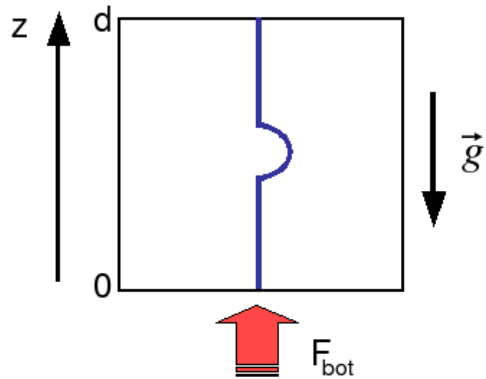}
 \caption{Scheme corresponding to our model. Gravity is pointing downward,
contrary to the vertical coordinate. The (blue) curve represents the
radiative conductivity profile that we are going to discuss further
whereas large (red) arrow expresses the radiative flux entering in the
bottom of the layer.}
 \label{fig:setup-intro}
\end{figure}

We are interested in small perturbations around an equilibrium state, that is, any
physical quantity is expanded around its mean value $X_0(\vec{r})$ as
\begin{equation}
 X(\vec{r},t)=X_0(\vec{r})+X'(\vec{r},t) \ltex{with} X'/X_0 \ll 1,
\end{equation}
where $X'$ is an Eulerian perturbation. Linearized continuity,
momentum and energy equations for the perturbations in a non-adiabatic
case are given by \citep[e.g.][]{Unno-book}

\begin{equation}
\left\lbrace
\begin{array}{lcl}
\lambda\rho' &=&\disp -\rho_0\Div\vu-\vu\cdot\na\ln \rho_0 \\ \\
\lambda\vu   &=&\disp -{1 \over \rho_0}\na p'+{\rho' \over\rho_0}\vg \\ \\
&& \disp +\nu\left(\Delta\vu+\frac{1}{3}\na\Div\vu+2\Vec{S}\cdot\na\ln\rho_0\right) \\ \\
\lambda T'&=&\disp -{1\over \rho_0 c_v}\Div\vec{F}'- (\gamma-1) T_0\Div\vu-\vu\cdot\na T_0 ,
\end{array}
\right.
\label{eq-tot1}
\end{equation}
where $\vu,\ \rho'\, \ T'$ and $\vec{F}'$  denote the velocity,
density, temperature and radiative flux perturbations, respectively.
The kinematic viscosity $\nu$ is supposed to be constant and $\Vec{S}$
denotes the traceless rate-of-strain tensor given by

\begin{equation}
S_{ij}=\dfrac{1}{2}\left(\dfrac{\partial u_i}{\partial x_j}+\dfrac{\partial u_j}{\partial x_i}-\dfrac{2}{3}\delta_{ij}\Div\vu\right) .
\label{stress-tensor}
\end{equation}
We seek normal modes with a time-dependence of the form $\exp(\lambda
t)$, where $\lambda=\tau + i\omega$. The real part $\tau$ denotes the
growing ($\tau >0$) or damping ($\tau<0$) rate whereas the imaginary
part $\omega$ denotes the frequency. We now assume that the layer is
fully radiative which, under the diffusion approximation, leads to the
following expression of the radiative flux perturbation

\begin{equation}
 \vec{F}'=-K_0\na T'-K' \na T_0,
\label{diffus}
\end{equation}
where $K_0$ denotes the radiative conductivity and $K'$ its Eulerian
perturbation. This perturbation $K'$ can be related in a general way to
the temperature one by

\begin{equation}
 \disp \frac{K'}{K_0}= \KT {T'\over T_0} 
\ltex{with} 
 \KT =\dfrac{d\ln K_0}{d \ln T_0}.
 \label{conductivity-perturbation}
\end{equation}
Finally, we impose the following boundary conditions

\begin{equation}
\left\lbrace
\begin{array}{l}
u_z=0 \ltex{for} z=[0,\ d], \\ \\ 
\dfrac{dT'}{dz}=0 \ltex{for} z=0 \ltex{and} T'=0 \ltex{for} z=d,
\end{array}
\right.
\label{boundary}
\end{equation}
which correspond to rigid walls at both limits of the domain, a perfect
conductor at the bottom and a perfect insulator at the top.

\section{Conditions for instability}
\label{ref:inst-cond}

\subsection{A first condition derived from the work integral}

We recall that the main aim of this work is to clarify the favourable
conditions which may sustain unstable radial modes in a plane parallel
layer. An advisable mean to study the physics of such instability
is the work integral. In the following, we are going to assume that
transformations are quasi-adiabatic. This approximation is sufficient in
the deeper layers of a star but becomes no longer valid near its surface where non-adiabatic effects dominate. Nevertheless, this approximation is useful when one wants to have an idea of the stability properties of an oscillation mode.

Using the work integral formalism in the quasi-adiabatic limit, we
demonstrate in Appendix \ref{appendix-work-integral} the following
expression for the damping or growing rate of an eigenmode

\begin{equation}
\tau =- \dfrac{\disp \Re \lc\int_V (\gamma -1) \dfrac{\delta \rho^*}{\rho}\Div \vec{F}' \dV\rc}
{\disp \int_V |\vec{u}|^2 \rho_0 \dV},
\end{equation}
where the symbol $\Re$ means the real part. We thus see that the sign
of $\tau$ only depends on the numerator which, under the diffusion
approximation, Eq.~(\ref{diffus}), leads to

\begin{equation}
 \tau \propto \Re\lc\disp\int_0^d\dfrac{\delta \rho^*}{\rho_0}\Div(K'\na T_0 + K_0 \na
T')\dz\rc.
\end{equation}
The main driving term in this expression is $K'\na T_0$ because it
represents the dynamical variation of opacity during an oscillation cycle,
which is the cause of $\kappa$-mechanism. We thus neglect $K_0\na T'$
to get

\begin{equation}
 \tau \propto \Re\lc \disp\int_0^d\dfrac{\delta \rho^*}{\rho_0}\Div(K' \na T_0)\dz\rc.
\end{equation}
As we can see in Fig.~\ref{fig:equilibrium-fields}b, the equilibrium
temperature is an almost linear function of $z$, except in the vicinity
of the conductivity hollow. Therefore, $\na T_0$ is almost constant thus we have
\begin{equation}
 \tau \propto \Re\lc \disp\int_0^d\dfrac{\delta \rho^*}{\rho_0}\na T_0\na K' \dz\rc.
\end{equation}
With Eq.~(\ref{conductivity-perturbation}), we have $K'=K_0(T_0)\KT
T'/T_0$.  As we are interested in low order modes we assume that $\na K'$
is dominated by $K_0(T_0) T'/T_0 \na \KT$. The same approximation  will
be done in Eq.~(\ref{approx1}). We then obtain

\begin{equation}
 \tau \propto \Re\lc\disp\int_0^d\dfrac{\delta \rho^*}{\rho_0}\na T_0 K_0(T_0)\dfrac{
T'}{T_0}\na \KT \dz\rc.
\end{equation}

Let us now consider a compression phase corresponding to $\delta \rho^*
/\rho_0 >0$ and $T'/ T_0 >0$. As $\na T_0 <0$, a necessary condition to
obtain unstable modes with $\tau > 0$ is then

\begin{equation}
 \dfrac{d\KT}{dz}<0.
 \label{criterion2}
\end{equation}
In variable stars, this condition can occur in ionisation regions. Indeed,
opacity tables clearly show that these regions are associated with
large ``bumps'' in opacity \citep[e.g. the review of][]{Carson,Seaton-Badnell}. With
Eq.~(\ref{criterion2}), one can obtain unstable modes if driving prevails
over damping. This result has the following physical meaning: if the
radiative conductivity decreases during a compression phase, then some
part of the radiative flux is blocked and some energy is stored during
compression contributing to increase the ratio of ionised matter.
During the decompression phase, this extra energy is released under
mechanical work to the environment and can excite the oscillations. It
works as Eddington predicted when he imagined a thermal heat engine to
justify Cepheids' oscillations \citep{Eddington}.

\subsection{A second condition derived from the so-called ``transition
region''}
\label{sec:trans}

In the previous paragraph, we have recalled a necessary condition
to obtain unstable modes (Eq.~\ref{criterion2}). Nevertheless,
the demonstration was made under the quasi-adiabatic approximation
which fails near the surface. As a consequence, we cannot know if the
driving caused by ionisation regions can prevail over other damping
regions. Indeed, ionisation regions where Eq.~(\ref{criterion2}) is
satisfied are very thin compared to the whole atmosphere of a star and
thus the influence of the damping regions on the instability remains questioning at this stage.
Hereafter, we will essentially follow Cox's demonstration
\citep[e.g.][]{Cox-book,Christensen}.

In Lagrangian variables, the energy equation can be written as

\beq
 \dnt{} \lp{\delta T\over T_0}\rp -(\gamma -1)\dnt{} \lp {\delta \rho \over \rho_0}\rp =
-\dfrac{\Fbot}{\rho_0 c_v T_0}\Div \left(\dfrac{\vec{F}'}{\Fbot}\right),
\eeq
where the symbol $\delta$ means Lagrangian perturbations. As we consider a 1-D box, we simply have

\beq
 -d\left(\dfrac{F'}{\Fbot}\right) dt = \dfrac{\rho_0 c_v T}{\Fbot} d\lc \dfrac{\delta
T}{T_0}-(\gamma-1)\dfrac{\delta \rho}{\rho_0}\rc\ \dz.
\eeq
Let us integrate between a given height $z$ and the surface to get

\beq
\Delta\left(\dfrac{F'}{\Fbot}\right) \Pi \sim {\langle c_v T_0 \rangle\over \Fbot} \int_z^d \rho_0
\hbox{dz'}\ d\lc\dfrac{\delta T}{T_0}-(\gamma-1)\dfrac{\delta \rho}{\rho_0}\rc,
\eeq
where $\Delta(F'/\Fbot)$ is the variation of $F'/\Fbot$ between the
considered point and the surface, $\langle c_v T_0 \rangle$ is an
average of $c_v T_0$ over this region and $\Pi$ is a characteristic
dynamical timescale (the pulsation period of the fundamental mode,
typically). As we are principally interested in the study of low order
modes (i.e. the fundamental one), we assume that the eigenvectors are
almost constant \citep[e.g.][]{Cox-book}. As a consequence, one obtains

\begin{equation}
 d\left[\dfrac{\delta T}{T_0}-(\gamma-1)\dfrac{\delta \rho}{\rho_0}\right] \approx
\left\langle\dfrac{\delta T}{T_0}-(\gamma-1)\dfrac{\delta \rho}{\rho_0} \right\rangle_z,
\label{approx1}
\end{equation}
where $\langle  . \rangle_z$ expresses an average of the eigenfunctions. Noting that

\beq
\int_z^d \rho_0 \hbox{dz'} = \dfrac{\Delta m}{S} \ltex{and} L=\Fbot\ S,
\eeq
leads to

\beq
 \Delta\left(\dfrac{F'}{\Fbot}\right) \sim \Psi \left\langle\dfrac{\delta
T}{T_0}-(\gamma-1)\dfrac{\delta \rho}{\rho_0}\right\rangle_z,
\label{eq-nrj-rescale}
\eeq
where
\begin{equation}
 \Psi=\dfrac{\langle c_v T_0\rangle \Delta m}{\Pi L}.
 \label{psi-param}
\end{equation}
$\Psi$ has the following physical meaning: it represents the ratio of
the thermal energy embedded between the considered point and the surface
to the energy radiated during an oscillation period\footnote{$\Psi$
can also be interpreted as the ratio of the local thermal timescale to
the dynamical one.}.  A second instability criterion can then be derived
from the value of this quantity $\Psi$:

\begin{itemize}

\item If the considered point is too deep in the star, the local
thermal timescale is longer than the dynamical one and $\Psi \gg 1$. As
a consequence, pulsation has no influence on the energy release and the
adiabatic approximation\footnote{One can  notice that the ``adiabatic''
approximation means here that deeply in the layer the mode has an
adiabatic behaviour whereas next to the surface it remains strongly
non-adiabatic.} is well-suited in this case.  

\item On the contrary, if the considered point is next to the surface,
then $\Delta m$ is very small and $\Psi \ll 1$. As a consequence,
the energy content (and the corresponding mass content) is too weak to
influence the radiative flux and the luminosity perturbation is said
to be \textit{frozen in}. It means that the radiative flux perturbation
doesn't vary anymore in this region.

\item Between these two regions, $\Psi$ can be $\od{1}$ which defines
a zone called the \textit{transition region} separating the adiabatic
interior from the strongly non-adiabatic surface, that is,

\begin{equation}
 \Psi=\dfrac{\langle c_v T_0 \rangle \Delta m}{\Pi L} \sim 1.
 \label{criterion}
\end{equation}
\end{itemize}
\citet{1968Cox} have first determined the temperature associated
with this transition region for different Cepheids' models. For the
fundamental mode, they have obtained $\Ttr \simeq 40\ 000$ K while
the transition lies nearer the surface for higher order modes. 

The corresponding positions of ionisation regions and transition
ones are therefore crucial for the instability \citep{Gilliland}. Indeed,
if they overlap -it corresponds to
the instability strip- the bottom of the ionisation region strongly
contributes to driving because it acts in a quasi-adiabatic place. On
the contrary, its top is in a strongly non-adiabatic region where the
luminosity is frozen in. As a consequence, the zone over ionisation region
is not damping and driving prevails in this case: modes become unstable
(see e.g. \cite{Cox-book} for a more detailed description).

We then sum up the conditions propitious to instability as
\beq
 \dfrac{d\KT}{dz} <0 \ltex{and} \Ttr\sim\Tion .
 \label{Cox-condition}
\eeq
With the conditions given in Eq.~(\ref{Cox-condition}), one
thus obtains $\Tion\simeq 40\ 000$ K. This temperature corresponds to
the second Helium ionisation which is known to be mainly responsible
for the driving of modes in Cepheids \citep[see][]{1963Cox}.

\section{Our hydrodynamical model}
\label{sec:hydro_model}

\subsection{The choice of the radiative conductivity profile}
\label{sec:choice}

\begin{figure}[htbp]
      \centering
      \includegraphics[width=9cm]{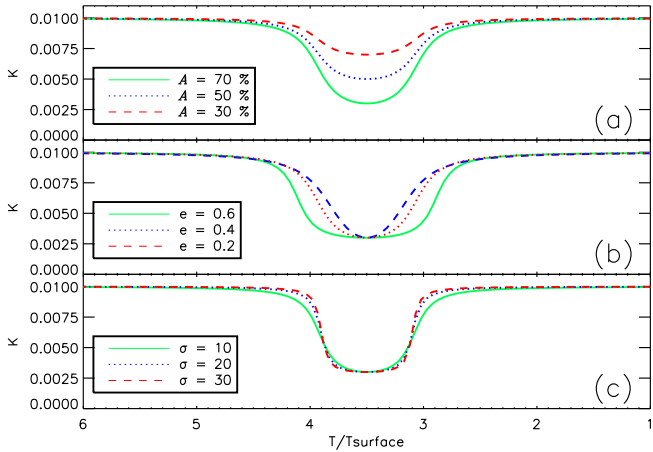}
      \caption{Influence of the hollow parameters on the conductivity profile for
      $\Kmax=10^{-2}$ and $\Tbump=3.5$:
      amplitude $\Amp$ (\textbf{a}), width $e$ (\textbf{b}) and slope $\sigma$
      (\textbf{c}) (here the abscissa denotes a
dimensionless temperature based on the surface one, i.e. T/T$_\text{surface}$)}.
      \label{fig:profile-rad}
\end{figure}

Strictly speaking, radiative conductivity $K_0$ depends both on temperature and density as diffusion approximation gives \citep[e.g.][]{Mihalas-Mihalas}

\begin{equation}
 K_0=\dfrac{16\tilde{\sigma}T^3}{3\kappa\rho},
\end{equation}
where $\tilde{\sigma}$ denotes Stefan-Boltzmann's constant. Kramer's
laws constitute good approximations of opacity laws with

\begin{equation}
\kappa \propto \rho^n T^{-s},
\end{equation}
where, for example, $n=1$ and $s=3.5$ for the free-free opacity
\citep[e.g.][]{Han94}. As shown in the Introduction and in
\S\ref{hydro_eqs}, we simply consider here a constant radiative
conductivity on which a hollow representing a bump in opacity is added.
This simple approach is an easy way to reproduce the physical conditions
propitious to $\kappa$-mechanism while keeping the ability to quickly
change the hollow amplitude, slope or width and achieving so an efficient
parametric study of the instability. As a consequence, the radiative
conductivity is given by\footnote{This $\arctan$-profile may appear
quite intricate, compared e.g. to a gaussian-like one, but it allows
us to change almost independently the hollow parameters (i.e. changing
its amplitude while keeping its width).}

\begin{equation}
 K_0(T)=\Kmax\left[1+\Amp\dfrac{-\pi/2+\arctan(\sigma T^+T^-)}{\pi/2+\arctan(\sigma e^2)}\right],
\label{conductivity-profile1}
\end{equation}
with
\begin{equation}
 \Amp=\dfrac{\Kmax-\Kmin}{\Kmax},\quad T^{\pm}=T-\Tbump \pm e,
\label{conductivity-profile2}
\end{equation}
and
\begin{itemize}
 \item $ \Tbump$ is the position of the hollow in temperature.
 \item $\Kmax$ and $\Kmin$ are the conductivity extrema, $\Amp$ being the corresponding relative amplitude.
 \item $\sigma$ represents the slope of the hollow.
 \item $e$ represents the half of the full width at half maximum (i.e. $\mbox{FWHM}/2$) of the hollow. 
\end{itemize}
Examples of common values of these parameters are given in
Fig.~\ref{fig:profile-rad}.

\subsection{The equilibrium setup}

Hydrostatic and radiative equilibria in the diffusion limit are given by

\begin{equation}
\left\lbrace\begin{array}{l}
\na p_0=\rho_0 \vg \\ \\
\Div \lc K_0(T_0) \na T_0 \rc =0 .
\end{array}\right.
\end{equation}
Assuming a constant radiative flux $\Fbot$ at the bottom of the layer leads to

\begin{equation}
\left\lbrace\begin{array}{l}
\disp \dnz{p_0} = -\rho_0 g \\ \\
\disp \dnz{T_0} = -{\Fbot \over K_0}.
\end{array}\right.
\end{equation}
Let us choose the depth of the layer $d$ as the length scale and the
temperature at the top of the layer $\Ttop$ as the temperature scale,
i.e $[z]=d$ and $[T]=\Ttop$. Moreover, the top density is chosen as the
density scale ($[\rho]=\rhotop$), velocity is given in units of $\sqrt{c_p
\Ttop}$, gravity in units of $c_p \Ttop /d$, pressure in units of $\rhotop
[u^2]$, while radiative conductivity is given in units of $\rhotop c_p d
[u]$. Corresponding radiative flux unit is then $\rhotop(c_p\Ttop)^{3/2}
$. The dimensionless equations then become

\begin{equation}
\left\lbrace\begin{array}{l}
\disp \dnz{\ln \tilde{p}_0} = -\dfrac{\gamma}{\gamma -1}{\tilde{g}\over \tilde{T}_0}  \\  \\
\disp \dnz{\tilde{T}_0} = -\dfrac{\tilde{F}_{\hbox{\scriptsize bot}}}{\tilde{K}_0} \\ \\
\disp \tilde{p}_0=\dfrac{\gamma -1}{\gamma}\tilde{\rho}_0\tilde{T}_0 .
\end{array}\right.
\label{eq-equil}
\end{equation}
The set of equations (\ref{eq-equil}) can be written in matrix form as

\begin{equation}
 A\vpsi =B(\vpsi),
\label{pb_equil}
\end{equation}
where $\vpsi =(p_0,T_0)^T$ is the equilibrium field vector and $A,
B$ are differential operators. One notes that the RHS, $B(\vpsi)$,
depends on the eigenvector itself through the terms $1/\tilde{T}_0$
and $1/\tilde{K}_0(\tilde{T}_0)$, which means solving a {\it nonlinear}
problem.

Finally, tildes on equilibrium fields emphasise dimensionless quantities 
but they will be dropped for clarity in the following. 

\subsection{About Schwartzschild's criterion in our model}

It is important to keep our box entirely radiative. That is why, we must take care of Schwartzschild's criterion \citep[e.g.][]{Chandra}

\begin{equation}
\left |\dfrac{d T_0}{dz} \right | < \left |\dfrac{d T_0}{dz} \right |_{\hbox{\scriptsize
adia}} \ltex{with} \left |\dfrac{d T_0}{dz} \right |_{\hbox{\scriptsize adia}} = \dfrac{g}{c_p}.
\end{equation}
If this inequality is ensured, then the layer is fully radiative. In our
dimensionless units, this condition becomes

\begin{equation}
 \left |\dfrac{dT_0}{dz} \right | < g,
 \label{Schwartz_criterion_final}
\end{equation}
meaning that large variations in temperature through the domain necessarily require large
values for the dimensionless gravity field $g$.

\subsection{Back to the radial oscillations equations}

To simplify the general system of oscillations equations (\ref{eq-tot1})
we first define the following new variables

\beq
R\equiv \rho'/\rho_0 \ltex{and} \theta\equiv T'/T_0.
\eeq 
Then, by using Eq.~(\ref{eq-equil}) and substituting the conductivity
equation into the energy one (see Appendix \ref{appendix-B}), one gets

\begin{equation}
\left\lbrace\begin{array}{lcl}
\lambda R&=&-\Div \vu-\vu\cdot\na\ln\rho_0 \\ \\
\lambda \vec{u} &=&-\dfrac{p_0}{\rho_0}\na (\theta+R)-\theta\vg \\ \\
&&+ \nu\left(\Delta\vu+\dfrac{1}{3}\na\Div\vu+2\Vec{S}\cdot\na\ln\rho_0\right) \\ \\
\lambda\theta &=&\dfrac{1}{\rho_0 c_v T_0}\Delta_z (K_0 T_0 \theta)
-(\gamma-1)\Div\vu-\vu\cdot\na\ln T_0,
\end{array}\right.
\end{equation}
We then adopt the same dimensionless quantities used in the equilibrium
equations (\ref{eq-equil})

\begin{equation}
\left\lbrace\begin{array}{lcl}
\lambda R &=& \disp -\dnz{u_z} -\dnz{\ln \rho_0}u_z \\ \\
\lambda u_z &=& \disp -\frac{\gamma -1}{\gamma}T_0 \left(\dnz{R}+\dnz{\theta}\right)+
g\theta + {\cal D}_{\nu} \\ \\
\lambda \theta &=& \disp \gamma\chi_0 \lc \ddnz{\theta}+2\lp \dnz{\ln K_0} +\dnz{\ln T_0}\rp\dnz{\theta} \right .\\ \\
&&+ \disp \left . \dnz{\ln T_0}\dnz{\KT}\theta\rc \disp -(\gamma - 1)\dnz{u_z}-\dnz{\ln T_0}u_z,
\label{eq-complet}
\end{array}\right.
\end{equation}
where $u_z$ denotes the vertical velocity, $\chi_0 =K_0/\rho_0 c_p$ the
radiative diffusivity and the viscous dissipative term ${\cal D}_{\nu}$
is given by

\begin{equation}
{\cal D}_{\nu} = \frac{4}{3}\nu\left( \frac{d^2 u_z}{dz^2}+\frac{d\ln \rho_0}{dz}
\frac{du_z}{dz}\right).
\end{equation}
This system (\ref{eq-complet}) may formally be written as a generalised
eigenvalue problem

\beq
A\vpsi =\lambda B \vpsi,
\label{pb_oscill}
\eeq
where $\lambda=\tau + \rm{i}\omega$ is the complex eigenvalue associated
with the eigenvector $\vpsi =(R,u_z,\theta)^T$ while $A,\ B$ denote
differential operators.

Finally the set of boundary conditions (\ref{boundary}) written for $(u_z,\ \theta)$ is
\begin{equation}
\left\lbrace\begin{array}{l}
u_z=0 \ltex{for} z=[0,\ 1] \\ \\
\disp\dnz{\theta}+\dnz{\ln T_0}\theta=0 \ltex{for} z=0 \\ \\
\theta=0 \ltex{for} z=1.
\end{array}
\right.
\label{pb_BC}
\end{equation}

\subsection{The numerical methods}

We solve the two linear algebra problems (\ref{pb_equil}) and
(\ref{pb_oscill}-\ref{pb_BC}) using the LSB code \citep[Linear Solver
Builder,][]{LSB}. Both problems are discretized on the Gauss-Lobatto grid
associated with Chebyshev's polynomials leading to two distinct numerical
problems:

\begin{enumerate}

\item {\it the equilibrium model}: the computation of the equilibrium
structure requires to solve a {\it nonlinear} problem. One way to do
that is to use the so-called ``Picard's method'' based on the fixed-point
algorithm \citep{Hairer,Fukushima}.  It consists in solving our set of
first-order ordinary differential equations by successive iterations,
that is, we advance

\beq
A\vpsi_{n+1} = B(\vpsi_n).
\eeq
This scheme converges quite well provided that the initial guess $\vpsi_0$
is not ``too far'' from the solution and that nonlinearities are weak.

\item {\it the eigenmodes}: we can compute the whole spectrum of complex
eigenvalues $\lambda$ using the QZ algorithm \citep{Moler} or just
compute a pair $(\lambda,\ \vpsi)$ around a given guess of $\lambda$
using the iterative Arnoldi-Chebyshev algorithm \citep{Arnoldi, Saad}.

\end{enumerate}

\section{Results}
\label{sec:results}

In Eq.~(\ref{Schwartz_criterion_final}), we have shown that the
temperature contrast across the layer (and also the associated density
and pressure ones) is limited by the gravity field $g$. It may however
be judicious not to restrict ourselves to small density contrasts as
the mass involved between the conductivity hollow and the surface,
that is $\Delta m$, enters in one of the two favourable criteria for
the instability, see Eq.~(\ref{criterion}). We will therefore consider
in the following a ``convenient'' value $g=7$.

Moreover, we have to keep as small as possible the radiative conductivity
as we want to avoid too large values of diffusivity $\chi_0$, that is,
we do not want to deviate too much from conditions of the applying of
quasi-adiabatic developments. We therefore choose $\Kmax=10^{-2}$.  As a
consequence, the system (\ref{eq-equil}) combined with Schwartzschild's
criterion (\ref{Schwartz_criterion_final}) restrict the possible values
of the imposed bottom flux and we thus set $\Fbot=2\times 10^{-2}$ in
the following. Using mild spatial resolutions, we choose a conservative
value for the kinematic viscosity, that is, $\nu=5\times10^{-4}$.

\begin{figure}[htbp]
 \centering
 \includegraphics[width=9 cm]{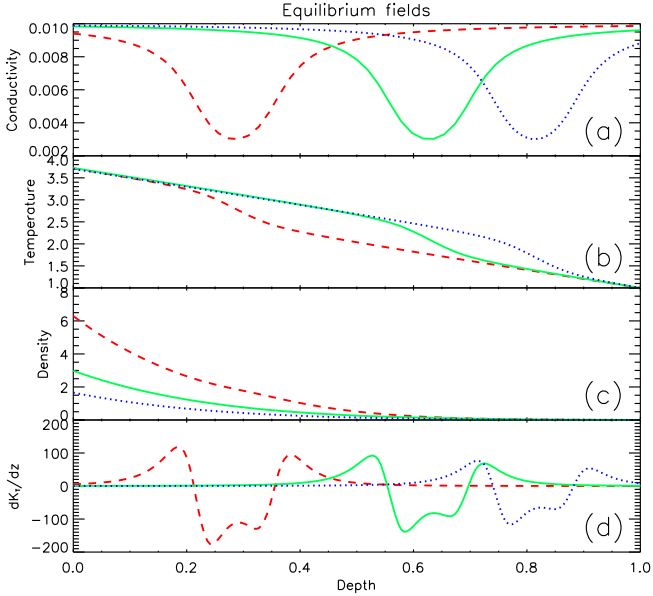}
 \caption{\textbf{a)} Three different conductivity hollows: $\Tbump=2.8$
 (dashed red line), $\Tbump=2.1$ (solid green line) and $\Tbump=1.7$
 (dotted blue line), with $\Amp=70\%,\ e=0.4$ and $\sigma=7$; \textbf{b)}
 corresponding temperature profiles; \textbf{c)} density profiles; \textbf{d)} 
 equilibrium profiles for $d \KT /dz$.}
 \label{fig:equilibrium-fields}
\end{figure}

\subsection{Computation of equilibrium fields}

To compute the equilibrium fields from second-order system
(\ref{eq-equil}), we must choose two different boundary conditions, one on
temperature and one on pressure. Without loss of generality, we first set
$T_0=1$ at the top of the layer. Second, as we are interested in having
the transition region of the fundamental mode located in the layer middle,
Eq.~(\ref{criterion}) has to be satisfied at that place. We have already
said that $\Fbot=2\times 10^{-2}$ and $\Kmax=10^{-2}$. Temperature value
is fixed by $\Fbot$, $\Kmax$ and its boundary condition. The pulsation
period of the fundamental mode is roughly expressed by

\begin{equation}
 \Pi\simeq \dfrac{2 d}{\langle c_s \rangle},
\end{equation}
where $\langle c_s \rangle$, denoting the mean sound speed, is related
to the temperature contrast across the layer. Thus, the only parameter
that we can change in Eq.~(\ref{criterion}) is $\Delta m$. This value is
directly linked to the boundary condition on density. In order to have
a suitable $\Delta m$-value in the box middle, we therefore decide to
take $\rho_0=2.5\times 10^{-3}$ at the top of the layer (corresponding
to $\ln P_0 =-6.9$).

Once chosen the values of the hollow parameters -$\Tbump, \Amp,\ e$ and 
$\sigma$-
we then solve the problem (\ref{eq-equil}) on the Gauss-Lobatto grid
to obtain the three equilibrium fields $T_0,\ P_0$ and $\rho_0$. With
these fields, it is possible to compute any physical quantities needed
in the oscillations equations (\ref{eq-complet}), e.g. $\KT$ or the
diffusivity $\chi_0$. Some examples of obtained equilibrium fields are
given in Fig.~\ref{fig:equilibrium-fields} for three different positions
of the conductivity hollow.

\subsection{Computation of eigenmodes}

\begin{figure}[htbp]
 \centering
 \psfrag{toto}[c][c]{Fundamental mode with $\omega_{00}=5.44$ and $\tau=2.05\times 10^{-2}$}
 \includegraphics[width=9 cm]{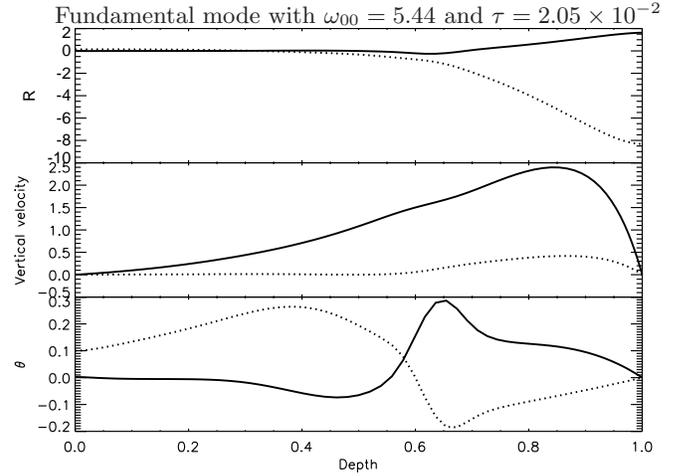}
 \caption{Eigenfunctions $(R,\ u_z,\ \theta)$ for the unstable fundamental
 mode, corresponding to a hollow with $\Tbump=2.1,\ \Amp=70\%,\ e=0.4$ and $\sigma=7$. The
 equilibrium setup used to compute this mode is the one displayed by a
 solid green line in Fig.~\ref{fig:equilibrium-fields}.}
 \label{fig:eigenvectors}
\end{figure}

With the different equilibrium quantities, we solve the eigenvalue
problem (\ref{eq-complet}) to get the whole spectrum of eigenvalues
$\lambda$. Among them, it is possible to obtain the nearest eigenvalue
from a guessed one using the iterative Arnoldi-Chebyshev algorithm. As
an example, Fig.~\ref{fig:eigenvectors} brings out the eigenfunctions
associated with the unstable fundamental mode.

\begin{figure}[htbp]
 \centering
 \includegraphics[width=9 cm]{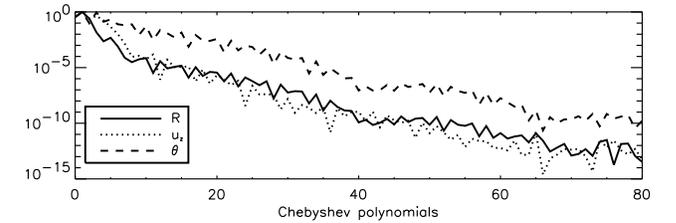}
 \caption{Chebyshev's spectra of the different fields $(R,\ u_z,\ \theta)$ for the
 eigenmode shown in Fig.~\ref{fig:eigenvectors}. Spectral precision is reached for every
 field.}
 \label{fig:spectrum}
\end{figure}

In order to check the convergence of this mode, we also compute
the spectra corresponding to the different eigenfunctions, see
Fig.~\ref{fig:spectrum}.

\begin{figure}[htbp]
 \centering
 \includegraphics[width=9 cm]{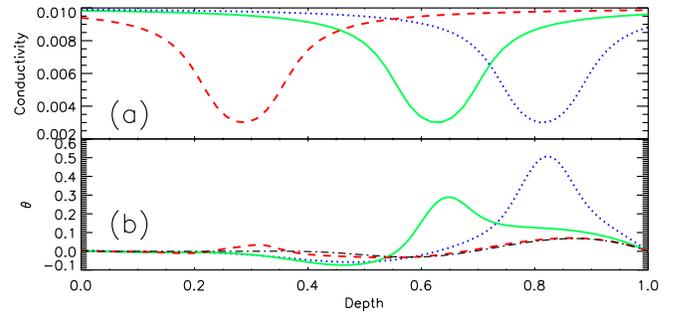}
 \caption{\textbf{a})Three different conductivity profiles with
 $\Tbump=1.7$ (dotted blue line), $\Tbump=2.1$ (solid green line) and
 $\Tbump=2.8$ (dashed red line); \textbf{b}) corresponding fundamental
 eigenmodes (here the real part of temperature perturbations
 $\theta$). The case with a constant radiative conductivity
 $K_0(T_0)=\Kmax$ is superimposed as a dot-dashed black line.}
 \label{fig:comp-temp}
\end{figure}

Finally, in order to determine the influence of the conductivity
hollow on that fundamental eigenvector, we compute three
eigenmodes with the different values of $\Tbump$ previously used
in Fig.~\ref{fig:equilibrium-fields}. The result is displayed in
Fig.~\ref{fig:comp-temp} for the temperature perturbation $\theta$, where
the case of a constant radiative conductivity profile (i.e. without a
conductivity hollow) is superimposed:

\begin{itemize}
\item The first case, corresponding to the dotted (blue) line, denotes
a hot star where the ionisation region is close to the surface. The
conductivity hollow has an influence on the temperature eigenfunction
which is strongly deformed at its location. We will show in
\S\ref{sec:work} that this deformation is not sufficient to destabilise
this mode as the $\Psi$-criterion should also be taken
into account (Eq.~\ref{criterion}).

\item The second one, corresponding to the solid (green) line, emphasises
a case where ionisation occurs approximately in the middle of the
layer. The eigenvector is still deformed by the conductivity hollow.

\item The last case, corresponding to the dashed (red) line, denotes
a cold star where ionisation region is far from the surface. For this
hollow location, the dynamical timescale is very small compared to the
local thermal one. As a consequence, thermodynamic transformations
are quasi-adiabatic and non-adiabatic effects are negligible
\textit{there}. This
is confirmed in Fig.~\ref{fig:comp-temp}b where the eigenmode is
practically not deformed by the conductivity hollow. To enlighten this
result, we have superimposed the case corresponding to a constant
radiative conductivity ($K_0(T_0)=\Kmax$). The obtained temperature
perturbations (the dot-dashed (black) line) are really next to the dashed
(red) one, meaning that the hollow has a marginal effect
on the mode stability.

\end{itemize}
We then show that the conductivity hollow has only an influence on the
shape of the eigenmode in upper parts of our layer. In deeper regions,
adiabatic thermodynamic transformations are overwhelming and the
eigenvector is pretty unchanged by the variations in the conductivity
profile.

\subsection{Parametric surveys of the instability}

As claimed in \S\ref{sec:choice}, our conductivity profile is well-suited
to deal with a parametric study of the $\kappa$-mechanism, as its
parameters $\Tbump,\ \Amp,\ e$ and $\sigma$ can easily be changed. We
thus next introduce the three parametric surveys which allowed us to
find the instability strips associated with the $\kappa$-mechanism.

\begin{figure}[t!]
 \centering
 \includegraphics[width=9 cm]{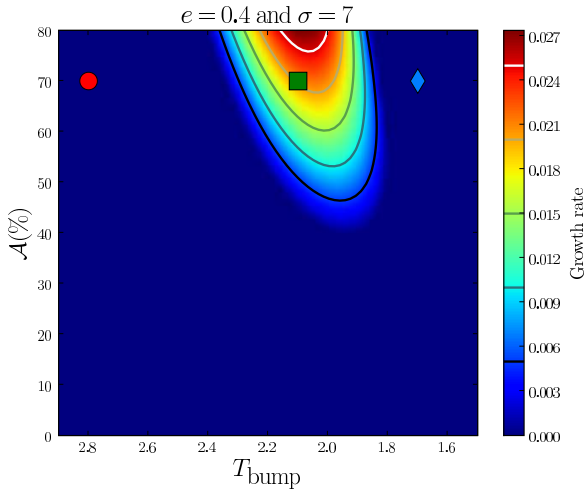}
 \caption{Instability strip for the fundamental mode in the plane
 $(\Tbump,\ \Amp)$ for given values of $e$ and $\sigma$ (isocontours in
 growth rates $\tau$ are displayed). The three marks correspond to the
 three particular computations done in Figs.~\ref{fig:equilibrium-fields}
 and \ref{fig:comp-temp} for $\Tbump=1.7$ (blue diamond, stable),
 $\Tbump=2.1$ (green square, unstable) and $\Tbump=2.8$ (red circle,
 stable).} \label{fig:ampli-Tbump}
\end{figure}

\subsubsection{The $\Tbump - \Amp$ survey}

At first, we want to determine the influence of the hollow amplitude on
stability. As a consequence, we choose a value for $\sigma$ and $e$ and
make $\Tbump$ and $\Amp$ vary. For each value of these two parameters,
we compute first the equilibrium fields and then, the eigenvalues with
their corresponding eigenvectors. Unstable modes are extracted among all
eigenvalues as their growth rate are positive. Fig.~\ref{fig:ampli-Tbump}
displays the obtained instability strip:

\begin{itemize}

\item Dark (blue) areas correspond to stable fundamental modes (i.e. with $\tau <0$).

\item Coloured areas correspond to unstable fundamental modes with the
lighter the colour the bigger the growth rate.

\end{itemize}

Two major results can then be derived from this figure: \textit{(i)}
a particular region in our layer ($\Tbump\in [1.8,\ 2.3]$) seems to be
propitious to the appearance of unstable modes, that is, one recovers the
concept of transition region discussed in \S\ref{sec:trans}; \textit{(ii)}
a minimal amplitude in the hollow is required to get an instability
($\Amp_{\hbox{\scriptsize min}} \simeq 45\%$).

\subsubsection{The $\Tbump - \sigma$ survey}

\begin{figure}[t!]
 \centering
 \includegraphics[width=9 cm]{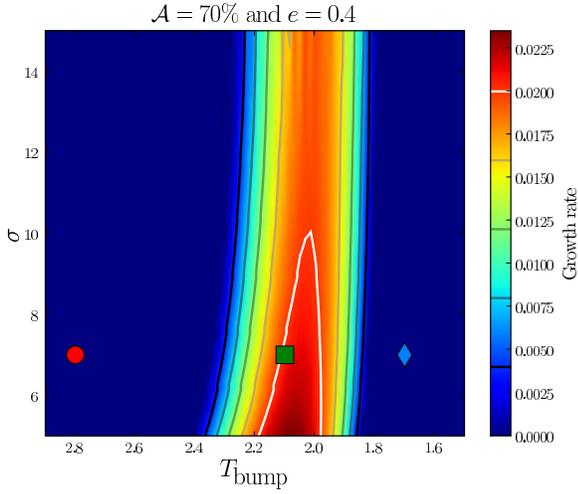}
 \caption{Instability strip for the fundamental mode in the plane $(\Tbump,\ \sigma)$,
 for given values of $\Amp$ and $e$. The blue diamond, green square and red circle
 correspond to the same modes displayed in Fig.~\ref{fig:ampli-Tbump}.}
 \label{fig:sigma-Tbump}
\end{figure}

Next, we study the influence of the slope $\sigma$ of the conductivity
hollow on stability. We thus choose a value for the amplitude $\Amp$ and
width $e$ while making $\Tbump$ and $\sigma$ vary. As for the previous
survey, we plot in Fig.~\ref{fig:sigma-Tbump} the isocontours in growth
rates $\tau$ but now in the plane $(\Tbump,\ \sigma)$. 

We then found the same kind of areas than in Fig.~\ref{fig:ampli-Tbump},
that is, an instability strip where modes are unstable (e.g. the
coloured region) embedded in large (dark) regions where modes are
stable. Nevertheless, the influence of the slope $\sigma$ on stability is
less significant than the amplitude one. In fact, the instability strip
covers the same temperature range than in Fig.\ref{fig:ampli-Tbump},
$\Tbump\in [1.8,\ 2.3]$, but it becomes almost vertical as there is no
critical value of $\sigma$ to trigger off the destabilisation of the
layer. In other words, there is a degeneracy in $\sigma$ as this parameter
is not affecting on the stability.

\subsubsection{The $\Tbump - e$ survey}

\begin{figure}[htpb]
 \centering
 \includegraphics[width=9 cm]{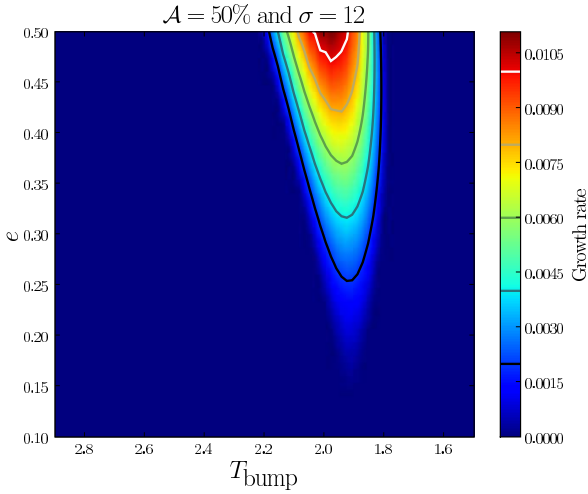}
 \caption{Instability strip for the fundamental mode in the plane $(\Tbump,\ e)$, for given
 values of $\Amp$ and $\sigma$.}
 \label{fig:ecart-Tbump}
\end{figure}

Finally, we study the influence of the hollow width $e$ on the stability
by performing a survey in the $(\Tbump,\ e)$-plane, while keeping constant
$\Amp$ and $\sigma$. Results are displayed in Fig.~\ref{fig:ecart-Tbump}.

An instability strip of the same kind than the one shown in
Fig.~\ref{fig:ampli-Tbump} is clearly visible as it exists a minimal
value for the width $e$ from which the fundamental mode becomes unstable
($e_{\hbox{\scriptsize min}}\simeq 0.15$). It means that narrow hollows
are not sufficient to initiate the instability.

\subsubsection{Summary of the surveys}

These three parametric studies allow us to show the respective influence
of the hollow parameters on the layer stability:

\begin{itemize}

\item In Figures \ref{fig:ampli-Tbump}-\ref{fig:ecart-Tbump},
we found a particular range in $\Tbump$, i.e. $\Tbump\in [1.8,\ 2.3]$,
within which the fundamental mode is unstable. It defines an instability
strip in temperature and corresponds to the particular area in the
layer we introduced before as the transition region.

\item Only the amplitude and width of the hollow have an influence
on the stability as we found critical values for both of them,
i.e. $\Amp_{\hbox{\scriptsize min}}\simeq 45\%$ and $e_{\hbox{\scriptsize
min}}\simeq0.15$. This can be linked to the condition (\ref{criterion2})
which entails that a precise shape in the conductivity profile is needed
to get unstable modes. Moreover, the hollow slope $\sigma$ is not a 
key control parameter as it does not modify the shape of the instability strip while varying.

\end{itemize}

We are now going to clarify these propitious conditions thanks to the
work integral formalism.

\subsection{Work integral in the non-adiabatic case}
\label{sec:work}

\begin{figure}[htbp]
 \centering
 \includegraphics[width=9 cm]{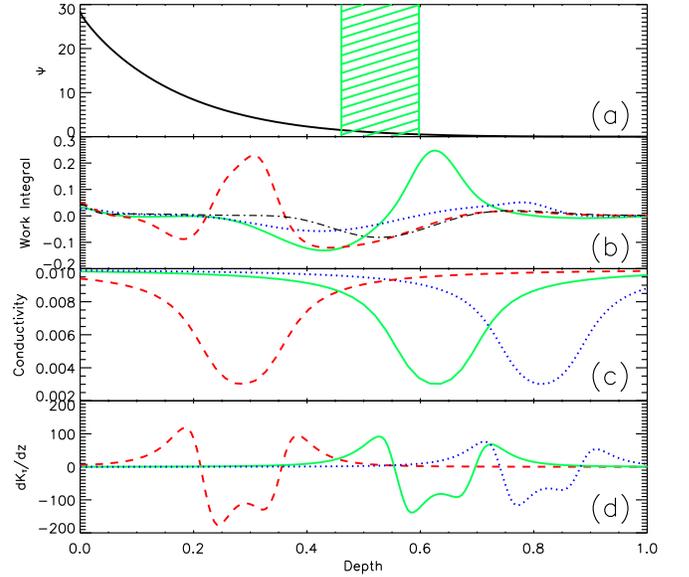}
 \caption{\textbf{a}) Coefficient $\Psi$ plotted over the entire box
 (Eq.~\ref{psi-param}). The green superimposed hatched zone represents
 the location where $\Psi = 1\pm 0.5$. \textbf{b}) The real part of
 the work integral numerator (Eq.~\ref{work-integral-final}) plotted
 for the three different equilibrium models already discussed in
 Figs.~\ref{fig:equilibrium-fields} and \ref{fig:comp-temp}. The
 case with a constant radiative conductivity is superimposed as a
 dot-dashed black line. \textbf{c}) Corresponding radiative conductivity
 profiles. \textbf{d}) Corresponding equilibrium field $d \KT /dz$.}
 \label{fig:wi2}
\end{figure}

By generalising to the non-adiabatic case the demonstration done in
Appendix \ref{appendix-work-integral}, we obtain the following expression
for the exact (complex) eigenvalue

\begin{equation}
\lambda=\dfrac{\displaystyle\int_0^1 \left[{\gamma-1\over\gamma}T_0(\theta+R)\dfrac{du_z^*}{dz}
-g R u_z^* +{\cal D}_{\nu} u_z^*\right ]\ \rho_0 \dz}{\displaystyle\int_0^1|u_z|^2\ \rho_0
\dz}.
\label{work-integral-final}
\end{equation}
As the denominator is always positive, the sign of the real part $\tau$
only depends on the numerator one. To separate the regions of damping
from the ones of driving, we thus plot the real part of the work integral
in Fig.~\ref{fig:wi2}b.

The work integral is really useful as it is possible to precisely check
where the driving occurs. The criteria given in Eq.~(\ref{Cox-condition})
predict that for a ``sufficient'' hollow, i.e. with a sufficient amplitude
and width located in the transition region, the fundamental mode
will be unstable. This result has already been checked thanks to the
parametric study (Fig.~\ref{fig:ampli-Tbump} for example). With the work
integral, we consider the same three particular modes studied before in
Figs.~\ref{fig:equilibrium-fields} and \ref{fig:ampli-Tbump}:

\begin{itemize}

\item The first one, plotted in dotted blue, expresses a case of a
hot star where ionisation region is next to the surface. In this case,
ionisation region is located in a place where density is really small,
and thus $\Psi \ll 1$. In Fig.~\ref{fig:wi2}b, we compare this case to
the one with a constant radiative conductivity, for which $K_O(T_0)=\Kmax$
and $d \KT /dz =0$, and we can conclude that the conductivity hollow has
a little influence on the work integral in this case: driving is unable
to prevail over damping and $\tau < 0$.

\item In the second case, plotted in solid green, the radiative
conductivity begins to decrease significantly at the location of the
transition region where $\Psi \simeq 1$. As a consequence, driving is
important in this place (Eq.~\ref{Cox-condition}). In addition, the
radiative conductivity increase occurs in a place where non-adiabatic
effects are already significant, i.e $\Psi < 1$. This means that no
damping will occur between the hollow position and the surface because
the radiative flux perturbations are ``frozen in''. Driving is overcoming
damping in this case and thus, $\tau > 0$.

\item The third case, plotted in dashed red, denotes a cold star where
ionisation region is located deeper in the stellar atmosphere where $\Psi
\gg 1$. As shown in Fig.~\ref{fig:comp-temp}, ionisation occurs there
in a quasi-adiabatic place. As a consequence, excitation provided by
the conductivity hollow cannot balance the damping arising in the layer,
thus $\tau <0$.

\end{itemize}

In conclusion, Fig.~\ref{fig:wi2} illustrates both conditions given
in Eq.~(\ref{Cox-condition}): \textit{(i)} it is first necessary to
have $d \KT /dz <0$ to drive the oscillations. But this condition is
not sufficient to sustain the $\kappa$-mechanism; \textit{(ii)} indeed,
the thermal engine underlying the conductivity hollow has to be located
neither too deep in the star, nor too close to the surface. Therefore,
only a specific range of effective temperatures allows the overlap of
the transition and ionisation regions, which leads to the instability
strips found in Figs.~\ref{fig:ampli-Tbump}-\ref{fig:ecart-Tbump}.

\section{Direct Numerical Simulations}
\label{sec:DNS}

In order to confirm all of the instability strips arising in the
previous linear stability analysis, we performed direct numerical
simulations of the \textit{nonlinear} problem. That is, starting
from the most favourable setups found during the parametric surveys,
we advanced in time the nonlinear hydrodynamic equations to
check:

\begin{itemize}

\item the onset of the instability sustained by the $\kappa$-mechanism
and thus confirm the growth rates of the linear stability analysis.

\item the nonlinear saturation of the instability, which is of
course not caught in the linear analysis, and have an estimation
of the final amplitudes of modes.

\end{itemize}

All DNS have been done using the Pencil Code\footnote{See
\href{http://www.nordita.org/software/pencil-code/}{http://www.nordita.org/software/pencil-code/} and
\cite{Pencil-Code}.}.
This non-conservative code is a high-order centered finite difference
code (of the sixth order in space and third order in time) and conserved
quantities are kept up to the discretization error of the scheme. On
multiprocessor computers, it uses the MPI libraries (Message Passing
Interface) which allow communications between processors and thus runs
in parallel. Moreover, this code is a fully explicit one as the
computation of the solution at time $t^{n+1}$ depends on the solution
obtained at time $t^n$ before. The timestep $dt$ is therefore limited
by an usual Courant-Friedrichs-Levy (CFL) condition based on the
consideration of the smallest physical timestep existing in the
simulation as

\begin{equation}
 dt \leq \min\left(c_{\delta t}\dfrac{\delta x}{c_{s_{\text{max}}}},c_{\delta t \chi}\dfrac{\delta x^2}{\chi_{\text{max}}},c_{\delta t \nu}\dfrac{\delta x^2}{\nu_{\text{max}}}\right),
\label{num:CFL-criterion}
\end{equation}
where  $c_{\delta t}$, $c_{\delta t \chi}$ and $c_{\delta t \nu}$
are constant coefficients depending on the spatial order of the
scheme \citep[see \S 19.2 in][hereafter PTVF92]{Numerical-Recipes}.

\subsection{The needed of an implicit solver}

\begin{figure}[htbp]
 \centering
 \includegraphics[width=9 cm]{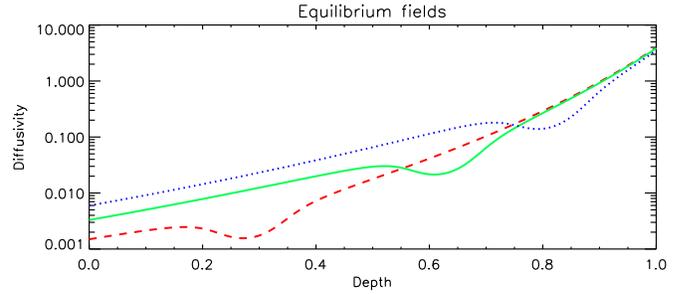}
 \caption{Three different diffusivity profiles with $\Tbump=1.7$ (dotted
blue line), $\Tbump=2.1$ (solid green line) and $\Tbump=2.8$ (dashed red
line).}
 \label{fig:chi-prof}
\end{figure}

This CFL condition is clearly a problem in our case as the most
favourable setups imply very large radiative diffusivities at the
surface. Fig.~\ref{fig:chi-prof} emphasises this point where the
diffusivity profile is plotted for the three common hollows used in
this work ($\Tbump=[1.7,\ 2.1,\ 2.8]$). Because the $\Psi$-criterion
(\ref{criterion}) imposes a weak value of $\Delta m$ and hence of
the top density, one gets large surface values for the radiative
diffusivity $\chi=K/\rho c_p\simeq 3$ in all cases. It means that the
corresponding radiative diffusion timestep $c_{\delta t \chi}\ \delta
x^2/\chi_{\text{max}}$ entering in Eq.~(\ref{num:CFL-criterion}) will
be the smallest one and will impose a very small timestep $dt$. As an
example, if we consider a typical spatial resolution of $256\times 256$
(i.e. 256 gridpoints in each direction), we obtain $dt \sim 10^{-6}$
whereas the dynamics of the layer is rather constrained by the sound speed
of which the dynamical timescale is $\delta x/c_{s_{\text{max}}}\simeq
3\times10^{-3}$.

It is therefore numerically prohibitive to reach the nonlinear
saturation of excited acoustic modes with such an explicit solver. To
limit the number of iteration, we have decided to solve the diffusion
of temperature \textit{implicitly}. In fact, as implicit schemes are
unconditionally stable, we have no more constraints on the timestep
coming from radiative diffusion and thus the CFL becomes

\begin{equation}
 dt \leq \min\left(c_{\delta t}\dfrac{\delta x}{c_{s_{\text{max}}}},c_{\delta t \nu}\dfrac{\delta x^2}{\nu_{\text{max}}}\right) ,
\end{equation}
giving $dt \sim 10^{-3}$ for the same resolution $256\times 256$.

\subsection{The Alternate Direction Implicit (ADI) scheme}

In order to solve the temperature equation, we adopt the time-split
formulation given in \citet{astro3d}. We thus solve first explicitly the
three hydrodynamic equations, i.e. density, velocity and temperature,
but without solving the radiative diffusion term at this step. We then
solve the temperature diffusion with the intermediate temperature
$T^{n+1/2}_{\text{expl}}$ treated as a source term.

The time advance of the diffusion temperature equation is treated implicitly in the form
\begin{equation}
 \dfrac{T^{n+1}-T^n}{dt}=\dfrac{T^{n+1/2}_{\text{expl}}
-T^n_{\text{expl}}}{dt}+\cal R ,
\label{eq:temp}
\end{equation}
where the radiative diffusion term $\cal R$ is expressed by
\begin{equation}
 \begin{array}{rcl}
\cal R&=& \dfrac{1}{2}\dfrac{1}{\rho^{n+1/2}_{\text{expl}} c_v}\Div 
\left[K(T^{n+1})\na T^{n+1}+K(T^n)\na T^n\right] \\ \\
 &=&\disp\frac{1}{2}[\Lambda_x(T^{n+1})+\Lambda_x(T^{n})+\Lambda_z(T^{n+1}
)+\Lambda_z(T^{n})] .
\end{array}
\end{equation}

$\cal R$ could be directly dealt with a Cranck-Nicholson method and a
single matrix inversion with, for instance, a Successive Over Relaxation
(SOR) method. For a 2-D problem, this approach forces us to invert a very
sparse $N^2\times N^2$ matrix (e.g. \S 19.5 in PTVF92). That is why we
have decided to implement an Alternate Direction Implicit (ADI) scheme
which relies on operator splitting theory which
is frequently used in diffusion problems \citep[e.g. \S 19.3 in
PTVF92; ][]{Dendy,Masalkar}. It allows
us to deal with tridiagonal matrices (or cyclic ones depending on
the boundary conditions used) by solving implicitly both directions
successively.

To treat implicitly the nonlinear terms (i.e. $K(T)$) we have adopted the
following approximation commonly called Rosenbrock's method
\citep{Rosenbrock, ADI-Witelski}
  
\begin{equation}
 \Lambda(T^{n+1})=\Lambda(T^n)+\underbrace{\left(\dfrac{\partial \Lambda}{\partial T^n}\right)}_{\disp\equiv J}(T^{n+1}-T^{n}) , \end{equation}
where $J$ denotes the Jacobian matrix associated with the operator 
$\Lambda$. Expanding Eq.~(\ref{eq:temp}) in each direction thus leads to

\begin{equation}
\left\lbrace
 \begin{array}{rcl}
 \left(I-\dfrac{dt}{2}J_x\right) \alpha &=& \Lambda_x(T^n)+\Lambda_z(T^n) + \dfrac{T^{n+1/2}_{\text{expl}}-T^n_{\text{expl}}}{dt} \\ \\
  \left(I-\dfrac{dt}{2}J_z\right) \beta &= &\alpha \\ \\
 T^{n+1}&=&T^{n}+dt\cdot\beta ,
\end{array}\right.
\label{num:schem-jacobian}
\end{equation}
where $I,\ J_x$ and $J_z$ are the identity matrix, a cyclic one and a
tridiagonal one, respectively.

\subsection{Results}

All 2D-simulations were carried out using a mean resolution $16\times
201$ and a constant kinematic viscosity $\nu = 5\times 10^{-4}$. In
order to avoid as much as possible the propagation of nonradial modes
\citep{Isa2007}, we chose a ``small'' box with an aspect ratio $L_x/L_z
= 1$. Indeed, if we refer to classical hydrodynamic instabilities,
the critical horizontal wavelength $\lambda_{\text{crit}}$
from which the instability develops is generally a bit larger
than the vertical extent $L_z$ of the domain. As an example,
$\lambda_{\text{crit}}/L_z=2\sqrt{2}=2.83$ in the Rayleigh-B\'enard
convection \citep{Chandra} or $\lambda_{\text{crit}}/L_z=2.84$ in the
compressible polytrope $m=1$ \citep{Gough-poly}.

\subsubsection{Growth rates from the DNS}

\begin{figure}[htbp]
 \centering
 \includegraphics[width=9 cm]{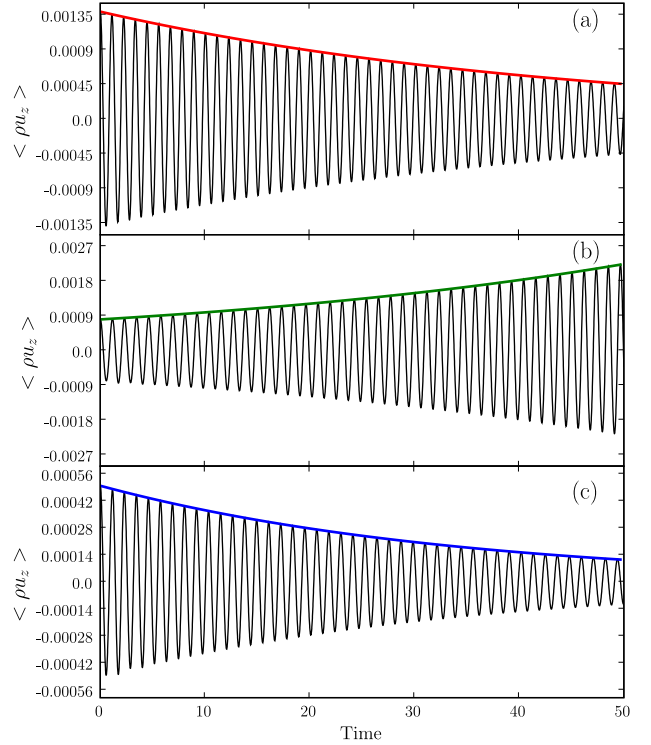}
 \caption{The temporal evolution for times $t \in[0,\ 50]$ of the mean
vertical momentum $\left<\rho u_z\right>$ with the three different
starting setups. In each panel, the growing (or damping) curve calculated
thanks to the linear stability analysis is superimposed. \textbf{a})
$\Tbump=2.8$ (red line). \textbf{b}) $\Tbump=2.1$ (green
line). \textbf{c}) $\Tbump=1.7$ (blue line). }
 \label{fig:comp-DNS}
\end{figure}

Fig.~\ref{fig:comp-DNS} emphasises the temporal evolution of the mean
momentum $\left<\rho u_z\right>$ (where $\left<.\right>$ is an average
over the entire box) for the three common equilibrium setups (the red,
green and blue ones corresponding to $\Tbump=[2.8,\ 1.7,\ 2.1]$). Only
one case appears to be unstable as:

\begin{table}[t]
\centering
\caption{Growth rates corresponding to the three equilibrium setups
given in Fig.~\ref{fig:comp-DNS} computed in the linear stability
analysis (the real part of eigenvalues) and in the DNS (computed from the exponential growth of the vertical momentum). The corresponding relative error is also given.}
\label{tab:growth-rates}
\begin{tabular}{cccc}
 &$\tau_{\text{LSB}}$ & $\tau_{\text{DNS}}$ & Rel. err. \\
 \hline \\
red & $-2.2415\times 10^{-2}$ &$-2.2412\times 10^{-2}$
& $1.5551\times 10^{-4}$ \\ \\
green & $\phantom{-}2.0484\times 10^{-2}$ & $\phantom{-}2.0516\times
10^{-2}$  & $1.5286\times 10^{-3}$ \\ \\
blue & $-2.9382\times 10^{-2}$ & $-2.9339\times 10^{-2}$  &
$1.4877\times 10^{-3}$
\end{tabular}
\end{table}

\begin{itemize}
\item The first (red) setup expresses the case of a cold star where
ionisation region is deep. The linear stability analysis achieved
before predicts that no excitation will occur in this case (the red
circle in Fig.~\ref{fig:ampli-Tbump} is well outside the instability
strip). As seen in Fig.~\ref{fig:comp-DNS}a, this result is confirmed
by the nonlinear simulation as the mean vertical momentum decreases
with time. Moreover, the damping rate calculated with LSB (superimposed
as a red line) is reproduced with a great agreement in this simulation
(Table~\ref{tab:growth-rates}).

\item The second (green) setup denotes the case where the
$\kappa$-mechanism is efficient. As a consequence, the fundamental p-mode
is expected to be unstable (the green square inside the instability
strip in Fig.~\ref{fig:ampli-Tbump}). The DNS confirms this excitation
(Fig.~\ref{fig:comp-DNS}b) and the growth rate is the same as the
predicted one (Table~\ref{tab:growth-rates}).

\item The third (blue) one corresponds to a hot star where ionisation
is next to the surface. In this case, damping phenomena prevail over
excitation and the fundamental mode is stable (the blue diamond outside
the instability strip in Fig.~\ref{fig:ampli-Tbump}). One more time,
the result is confirmed by the DNS (see Fig~.\ref{fig:comp-DNS}c and
Table~\ref{tab:growth-rates}).

\end{itemize}

\begin{figure}[htbp]
 \centering
 \includegraphics[width=9 cm]{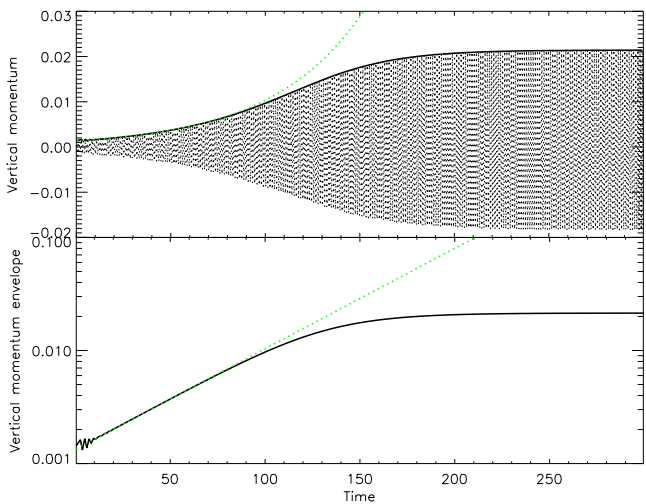}
 \caption{\textit{Upper panel}: temporal evolution of the mean vertical
 momentum $\left<\rho u_z\right>$. The linear stability analysis result
 from Fig.~\ref{fig:eigenvectors} is superimposed in green. \textit{Lower
 panel}: temporal evolution of the corresponding envelope with the
theoretical curve of growth still superimposed. Ordinate is scaled
logarithmically.}
 \label{fig:growth-rate}
\end{figure}

In Fig.~\ref{fig:growth-rate}, we have integrated the DNS from the initial
green setup with $\Tbump=2.1$ till the approach to the nonlinear limit cycle stability, that is, the nonlinear saturation. The
(green) dotted line corresponds to the theoretical growth rate given
in Table~\ref{tab:growth-rates}, that is, $\tau=2.0484\times 10^{-2}$.
The saturation of this mode appears to be achieved around time
$t\simeq 200$, which roughly corresponds to $170$ mode periods. Such
time interval is compatible with the characteristic timescale of the
instability given by $1/\tau \sim 100$. Finally, the amplitude reached
at the end of the saturation is about $\langle \rho u_z \rangle \simeq
2\times 10^{-2}$. A careful study of the modes saturation is
beyond the scope of this paper and will be developed in future works.

\subsubsection{Vertical profiles}

The good agreement between the theoretical and DNS growth rates shown in
Table~\ref{tab:growth-rates} marks a first success in reproducing the
$\kappa$-mechanism in our simulations. We next address the point of the 
modes structure by
computing the vertical profiles from the DNS and comparing them to the
eigenvectors of the linear stability analysis.

\begin{figure}[h!]
 \centering
 \includegraphics[width=9 cm]{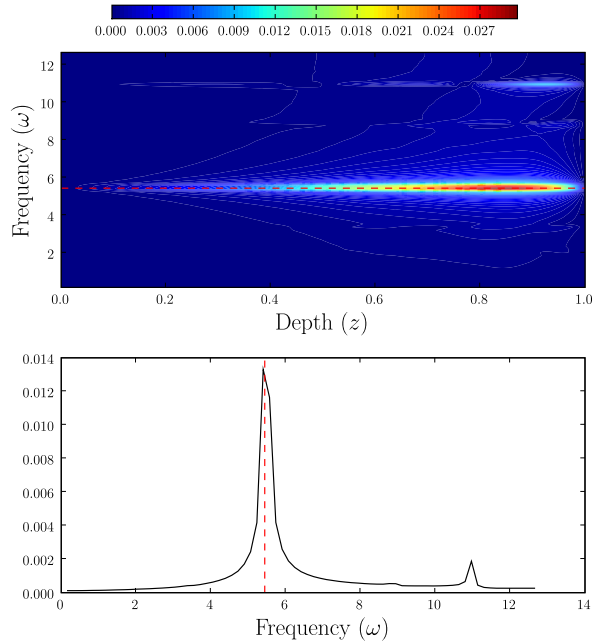}
 \caption{\textit{Upper panel}: temporal power spectrum in the
 $(z,\ \omega)$-plane for the radial modes $k_x=0$. \textit{Lower
 panel}: the resulting spectrum after an integration in
 depth. Dotted lines correspond to the theoretical frequency
 $\omega_{00} \simeq 5.44$ obtained in the linear stability analysis
 (e.g. Fig.~\ref{fig:eigenvectors}).}
 \label{fig:fourier}
\end{figure}

In Fig.~\ref{fig:fourier}, we have performed an horizontal and temporal
Fourier transforms of the vertical velocity field and plotted the
resulting power spectrum at $k_x=0$ in the plane $(z,\ \omega)$-plane.
With this method, we are able to determine exactly which acoustic
modes are excited or not in our numerical experiment as they emerge
in this plane as ``shark fin'' peaks around given frequencies
\citep[see][]{Boris-gmodes}. That is indeed what is displayed
in Fig.~\ref{fig:fourier} where the radial acoustic mode with a
frequency $\omega_{00}\simeq 5.44$ is excited and the agreement with
the linear eigenfrequency is remarkable. The second overtone with
$\omega_{02}\simeq 11.05$ also appears in the $(z,\ \omega)$-plane, with still
a good agreement between the linear stability analysis and the DNS. In
fact, one can note that this frequency is almost twice the fundamental
mode one. It means that a resonance-like interaction occurs between
these two modes: the $\kappa$-mechanism excites the fundamental mode and
a nonlinear interaction transfers some energy between this mode and the
second \textit{damped} overtone, leading to the nonlinear saturation.

\begin{figure}[htbp]
 \centering
 \includegraphics[width=9 cm]{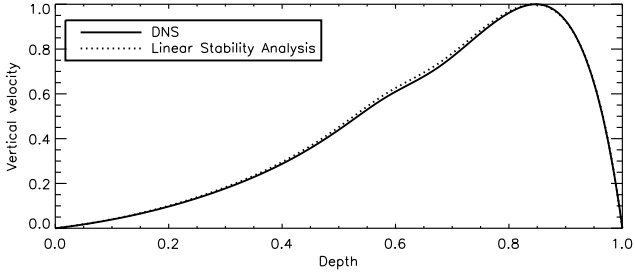}
 \caption{Comparison between normalised vertical velocity profiles
 $u_z$ deduced from the DNS power spectrum (solid line) and the linear
 stability analysis (dashed line).} 
\label{fig:vert-prof}
\end{figure}

As we have computed, thanks to LSB, the eigenfunctions from a
given equilibrium setup (see Fig.~\ref{fig:eigenvectors}), one
can compare them to the vertical velocity $u_z$ deduced from the
power spectra in Fig.~\ref{fig:fourier}. That is what is displayed
in Fig.~\ref{fig:vert-prof} where a mean profile around the frequency
$\omega_{00}=5.44$ has been computed. The DNS profile and the linear
stability analysis one perfectly overlap meaning that we both have an
agreement on the temporal (i.e. the frequency) and spatial (i.e. the
pattern) scales of this mode.

\section{Conclusion}
\label{sec:conclusion}

We have modeled $\kappa$-mechanism in Cepheids by a similar physical
problem: the propagation of radial acoustic waves in a partially
ionised shell. Our model consists in a perfect gas embedded in an
entirely radiative layer and the ionisation region has been depicted
by a configurable hollow in radiative conductivity. This approach
allows to quickly change the shape and location of this ionised
region in the layer through the hollow parameters $\Tbump,\ \Amp, \
\sigma$ and $e$
(Eqs.~\ref{conductivity-profile1}-\ref{conductivity-profile2}). We first
have checked that a hollow with a sufficient amplitude and width may
lead to

\begin{equation}
 \dfrac{d\KT}{dz} < 0,
\label{crit1}
\end{equation} 
which corresponds to the classical instability criterion derived from the
work integral and quasi-adiabatic considerations. Nevertheless, another
condition is needed to obtain a thermal heat engine in Eddington's sense:
the ionisation region has to be located at a certain place in the layer,
neither to deep nor to close to the surface. We have shown that this
propitious area is located in the so-called \textit{transition zone}
separating the quasi-adiabatic interior from the strongly non-adiabatic
surface. This thermodynamical criterion can be summarised by

\begin{equation} \Tion
\sim \Ttr  \ltex{and} \dfrac{\langle c_v \Ttr \rangle \Delta m}{\Pi L}
\sim 1.  
\label{crit2}
\end{equation} 

With this second criterion, appropriate boundary conditions have been
chosen in order to have a transition region in the middle of our box for
the fundamental acoustic mode. Both radiative and hydrostatic equilibria
have then been discretized on the (spectral) Gauss-Lobatto grid and
solutions have been computed thanks to a linear solver. Once done, we
have performed a linear stability analysis by computing the whole spectrum
as well as associated eigenvectors for the radial oscillations equations.

The main advantage of our approach lies in its ability to make
the parameters of the conductivity hollow vary. The major result of
parametric surveys is the checking of the previous conditions by way of
the appearing of instability strips: only configurations satisfying both
conditions (\ref{crit1}-\ref{crit2}) have led to unstable fundamental
modes (e.g.~Figs.~\ref{fig:ampli-Tbump} and \ref{fig:wi2}).

Then, these most favourable setups found in the linear stability
analysis have been the starting points of our first 2-D (nonlinear)
DNS of $\kappa$-mechanism. However, because of a too restrictive
Courant-Friedrichs-Levy (CFL) condition at the surface, these setups
were not well-suited for a fully explicit code such as the Pencil Code
one. We therefore have developed a new \textit{implicit} module to deal
with the radiative diffusion term and thus soften the CFL constraint.
These DNS have confirmed with a great agreement both growth rates and
structures of linearly unstable modes (see Figs.~\ref{fig:growth-rate}
and \ref{fig:vert-prof}). Moreover, we have been able to reach the
nonlinear saturation which involves an intricate coupling between the
fundamental mode and (at least) the damped second overtone of which the
period is a multiple of the fundamental one.

This work constitutes a first step in our Cepheids' project devoted to
the modelling of the convection-pulsation interaction in the coldest
Cepheids close to the red edge of the instability strip. Indeed,
it is well known that convection occurs in a non-negligible part
of these stars and modify the pulsation properties. For now,
models based on time-dependent convection theories reproduce quite well
the position of this red edge or, say, observational periods and light
curves \citep[YKB98;][]{Bono}. However, they involve many free parameters
which are either fitted to the observations or hardly constrained by
theoretical values, leading to almost similar results for different
parameter sets \citep{Koll02,Szabo07}.
 
Despite their own limitations (their weak contrasts in
pressure through the computational domain or their thermal
timescale problem, see e.g. \cite{Bran00}), DNS are without a doubt
a good way to address this convection-pulsation interaction as they
fully take into account the crucial nonlinearities. With a conductivity
profile solely based on temperature (Eq.~\ref{conductivity-profile1}),
it is easy to \textit{locally} shape one (or several) convective zone
by increasing the temperature gradient above the adiabatic one. Indeed,
as an ionisation region corresponds to a local increase in opacity,
it is well known that convection can develop there. This occurs in cold
Cepheids where two separate convectively unstable regions superimpose with
the HeI/H and HeII ionisation regions. By adjusting in DNS the convective
zone width or the strength of gravity, it will be possible to match the
(local) turnover timescale of convection with the mean mode period,
and then to study the coupling between convection and pulsation.

\begin{acknowledgements}
Calculations were carried out on the CalMip machine of the “Centre
Interuniversitaire de Calcul de Toulouse” (CICT) which is gratefully
acknowledged. It is also a pleasure to thank Isabelle Baraffe, Fabien
Dubuffet, Marie-Jo Goupil and Michel Rieutord for their fruitful
comments. 
\end{acknowledgements}

\begin{appendix}

\section{Work Integral}
\subsection{Using the quasi-adiabatic approximation}
\label{appendix-work-integral}

In the energy equation in Syst.~(\ref{eq-tot1}), we can first separate
adiabatic terms from non-adiabatic ones as

\begin{equation}
 T'=T'_{\hbox{\scriptsize adia}}-\dfrac{1}{\rho_0 c_v \lambda}\Div
\vec{F}',
\end{equation}
and, by using the ideal gas equation of state (\ref{EOS}),

\begin{equation}
 p'=p'_{\hbox{\scriptsize adia}}-\dfrac{\gamma-1}{\lambda}\Div \vec{F}'.
\end{equation}
This equation corresponds to the pressure perturbation due to both
adiabatic and non-adiabatic oscillations. The momentum equation in
Syst.~(\ref{eq-tot1}) then becomes

\begin{equation}\begin{array}{l}
\left(\lambda + \delta
\lambda\right)(\vu+\delta\vu)=\underbrace{-\dfrac{1}{\rho_0}\vec{\nabla}
p'_{\hbox{\scriptsize
adia}}+\dfrac{\rho'_{\hbox{\scriptsize
adia}}}{\rho_0}\vec{g}}_{\mbox{Adiabatic part}} \\ \\
\phantom{\left(\lambda + \delta\lambda\right)(\vu+\delta\vu)=}
\underbrace{+\dfrac{1}{\rho_0}\dfrac{\gamma
- 1}{\lambda}\vec{\nabla}\left(\Div
\vec{F}'\right)}_{\mbox{Non-adiabatic perturbation}}.
\end{array}
\end{equation}
This equation can formally be written as a generalised eigenvalue problem
with a perturbation operator

\begin{equation}
(A+\delta A)(\vpsi+\delta \vpsi)=(\lambda +\delta \lambda) B (\vpsi+\delta
\vpsi),
\end{equation}
where $\delta \lambda$ and $\delta \vpsi$ are caused by the perturbation
operator $\delta A$ (here the non-adiabatic effects).  $\delta\lambda$ can
then be obtained from a first-order perturbation analysis equivalent to
these done in quantum mechanics \citep[e.g.][]{Bender-Orzag,Dyson-Schutz}

\beq
\delta\lambda=\dfrac{\langle \vpsi_0 | \delta A \vpsi_0 \rangle}{\langle
\vpsi_0 | B \vpsi_0 \rangle},
\eeq
where the symbol $\langle\ |\ \rangle$ means the following dot product
\citep[e.g.][]{Lynden-Bell}

\beq
\langle \vf_1 |\vf_2 \rangle= \int_V \vf^*_1 \cdot \vf_2\ \rho_0 \rm{\dV}.
\eeq
We thus obtain the expression for the eigenvalue perturbation
$\delta \lambda$

\beq
\delta\lambda=\dfrac{\gamma - 1}{\lambda}\dfrac{\displaystyle\int_V
\vec{u^*}.\vec{\nabla}\left(\Div \vec{F}'\right)
\dV}{\displaystyle\int_V |\vec{u}|^2\rho_0 \dV}.
\eeq
We can go a step further using

\beq
\vec{u^*}.\vec{\nabla}\left(\Div \vec{F}'\right)=\Div\left(\vec{u^*}\Div
\vec{F}'\right) - \Div \vec{u^*}\Div
\vec{F}',
\label{eq_div}
\eeq
of which the first term in the RHS vanishes due to the rigid boundaries for the velocity
as

\begin{equation}
 \int_V\Div\left(\vec{u^*}\Div \vec{F}'\right) \dV=\oiint_S \vu^*\Div
\vec{F}'\cdot\vec{dS}=0.
\end{equation}
The second term in the RHS can also be simplified by using the continuity
equation

\begin{equation}
 \Div \vec{u^*}=-\lambda^*\dfrac{\delta \rho^*}{\rho_0},
\end{equation}
where $\delta\rho$ is the Lagrangian perturbation of density. We finally
get the expression for the eigenvalue perturbation in the non-adiabatic
case as

\begin{equation}
\delta\lambda=- \dfrac{\disp \int_V (\gamma -1) \dfrac{\delta
\rho^*}{\rho}\Div \vec{F}' \dV}
{\disp \int_V |\vec{u}|^2 \rho_0 \dV},
\label{app-result}
\end{equation}
where we have assumed that the adiabatic eigenmode is purely imaginary,
that is $\lambda={\rm i} \omega$,  and thus $\lambda^*/\lambda=-1$. We
note that $\Re(\delta \lambda)$ corresponds to the damping (or growing)
rate of a non-adiabatic mode, commonly written $\tau$.

\subsection{A thermodynamical approach}

It is also possible to derive Eq.~(\ref{app-result}) by energetic
considerations. Hereafter, we will essentially follow Hansen's demonstration
\citep{Han94}. Let us adopt a thin shell of mass $dm$ at a certain
radius. During a complete cycle of oscillations the work $dW$ done by
the shell on its surroundings is linked to the internal energy $dU$ and
heat $dq$ gained by this shell through the first principle of
thermodynamics 

\begin{equation}
 dq=dU+dW.
\end{equation} 
Integrating over a cycle of oscillations leads to

\begin{equation}
 \oint dq =\oint dU + \oint dW .
\end{equation}
We are now going to suppose that the oscillation mechanism is
quasi-adiabatic. It implies that every thin shell $dm$ behaves as a
Carnot-like heat engine where each process is reversible and the shell
comes back to its initial position after each cycle of oscillations.
Internal energy $dU$ being a state variable, one gets $\oint dU=0$
hence

\begin{equation}
 W=\oint dq.
\label{app-1}
\end{equation}
Now one applies the second principle of thermodynamics claiming that

\begin{equation}
 dS= \dfrac{dq}{T} .
\end{equation}
After some time elapsed, we have $T=T_0+\delta T$ and then

\begin{equation}
 dS=\dfrac{dq}{T_0}-\dfrac{\delta T}{T_0^2}dq,
\end{equation}
as a first-order approximation, with $\oint dS =0$ for the same reason
than $U$. We thus write

\begin{equation}
 \oint dq =\oint \dfrac{\delta T}{T_0}dq.
 \label{app-2}
\end{equation}
With Eqs.~(\ref{app-1}) and (\ref{app-2}), the work over a cycle is given by

\begin{equation}
 W=\oint \dfrac{\delta T}{T_0}dq,
\end{equation}
which leads to the total work integrated over mass shells

\begin{equation}
 W_{\hbox{\scriptsize tot}}=\displaystyle\int_M\oint \dfrac{\delta T}{T_0}dq dm.
 \label{app-3}
\end{equation}
If $W_{\hbox{\scriptsize tot}} >0$, the star produces work over a cycle
of oscillations and the initial perturbation will grow. In this case,
the star is driving pulsation which is thus unstable. It may occur when
shells gain heat (i.e. $dq >0$) during compression phases (i.e. $\delta
T >0$) and this is the so-called ``valve-mechanism'' proposed by
\citet{Eddington}.

Let us write the energy equation in the form

\begin{equation}
 dq=-\dfrac{1}{\rho_0} \Div \vec{F}' dt, 
\end{equation}
and substitute it in Eq.~(\ref{app-3})

\begin{equation}
 \dfrac{d W_{\hbox{\scriptsize tot}}}{ dt}= -\displaystyle\int_V \dfrac{\delta T}{T_0}\Div \vec{F}' \dV.
 \label{app-wtot}
\end{equation}
Kinetic energy over a period of oscillations is given by

\begin{equation}
 E=\dfrac{1}{2}\displaystyle\int_V |\vec{u}|^2 \rho_0 \dV,
 \label{app-NRJ}
\end{equation}
and Eqs.~(\ref{app-wtot}-\ref{app-NRJ}) lead to the following expression
for the growth rate of a mode

\begin{equation}
 \tau=\dfrac{1}{2}\dfrac{d W_{\hbox{\scriptsize tot}}/ d t}{E}.
\label{app-resNRJ}
\end{equation}
Note that the factor $1/2$ comes from the fact that the pulsation
amplitude grows or decays one half as does energy. The two formulations
Eqs.~(\ref{app-result}) and (\ref{app-resNRJ}) are of course alike by
assuming the quasi-adiabatic approximation $\delta T/T_0 \simeq (\gamma
-1)\delta \rho /\rho_0$.

\section{The radiative term in energy equation}
\label{appendix-B}

Radiative flux perturbations can be written as

\begin{equation}
 -\Div \vec{F}'=\Div\lp K_0\na T' +K'\na T_0 \rp.
\end{equation}
By assuming $\theta \equiv T'/T_0$ and using
Eq.~(\ref{conductivity-perturbation}), it entails

\begin{equation}
 -\dfrac{d}{dz} \vec{F}'=\Div\lp K_0
T_0\dfrac{d\theta}{dz}+K_0\theta\dfrac{d T_0}{d z}+
K_0\KT\theta\dfrac{dT_0}{dz} \rp\vec{e_z}.
\end{equation}
We then expand $\KT$ with equilibrium equations (\ref{eq-equil}) as

\begin{equation}
 \KT=\dfrac{d\ln K_0}{d \ln
T_0}=-\dfrac{T_0}{\Fbot}\dfrac{d K_0}{dz},
\end{equation}
and thus obtain

\begin{equation}
 -\Div\vec{F}'=\dfrac{d}{dz}\lp K_0
T_0\dfrac{d\theta}{dz}+K_0\theta\dfrac{d T_0}{d z}+\theta T_0 \dfrac{d
K_0}{dz} \rp.
\end{equation}
Finally, one gets

\begin{equation}
 -\Div\vec{F}'=\dfrac{d^2}{dz^2}\lp K_0 T_0\theta \rp=\Delta_z \lp K_0 T_0 \theta \rp.
\end{equation}

\end{appendix}

\end{document}